\documentclass[conference]{IEEEtran}
\IEEEoverridecommandlockouts
\usepackage{cite}
\usepackage{amsmath,amssymb,amsfonts}
\usepackage{algorithmic}
\usepackage{graphicx}
\usepackage{textcomp}
\usepackage{xcolor}
\usepackage{xspace}
\usepackage{tabularx}
\usepackage{multirow}
\usepackage{subfig}
\usepackage{todonotes}
\usepackage{enumitem}
\usepackage{siunitx}
\usepackage{authblk}
\usepackage{makecell}

\newtheorem{definition}{Definition}
\def\BibTeX{{\rm B\kern-.05em{\sc i\kern-.025em b}\kern-.08em
    T\kern-.1667em\lower.7ex\hbox{E}\kern-.125emX}}

\newcommand{\ourname}{\textsc{D\"IoT}\xspace}

\newcommand{\sgw}{\textit{Security Gateway}\xspace}
\newcommand{\iotss}{\textit{IoT Security Service}\xspace}

\newcommand{\anomalydetection}{\textit{Anomaly Detection}\xspace}
\newcommand{\ChallengeDynamic}{\textsf{C1}\xspace}
\newcommand{\ChallengeResourceLimit}{\textsf{C2}\xspace}
\newcommand{\ChallengeHeterogeneity}{\textsf{C3}\xspace}
\newcommand{\ChallengeScarcity}{\textsf{C4}\xspace}
\newcommand{\Activity}{\textit{Activity}\xspace}
\newcommand{\Deployment}{\textit{Deployment}\xspace}
\newcommand{\Attack}{\textit{Attack}\xspace}

\newcommand{\pkt}{\mathit{pkt}}
\newcommand{\sym}{s}
\newcommand{\prob}{p}

\newcommand{\devtype}{\mathit{type\#k}}

\newcommand{\map}{\mathit{mapping}_\devtype}
\newcommand{\win}{w}

\newcommand{\detThr}{\delta}
\newcommand{\anomalyThr}{\gamma}

\usepackage{xcolor}
\newcommand{\revision}[1]{{#1}}

\newcommand{\mcrot}[4]{\multicolumn{#1}{#2}{\rlap{\rotatebox{#3}{#4}~}}} 
\newcommand{\bul}{\textbullet}
\newcommand{\cir}{$\circ$}

\usepackage{fancyhdr}

\begin{document}

\title{\ourname: A Federated Self-learning \\Anomaly Detection System for IoT}


\author[1]{Thien Duc Nguyen}
\author[2]{Samuel Marchal}
\author[1]{Markus Miettinen}
\author[1]{Hossein Fereidooni}
\author[2]{\\ N. Asokan}
\author[1]{Ahmad-Reza Sadeghi}
\affil[1]{TU Darmstadt, Germany - \{ducthien.nguyen,markus.miettinen,hossein.fereidooni,ahmad.sadeghi\}@trust.tu-darmstadt.de}
\affil[2]{Aalto University, Finland - samuel.marchal@aalto.fi, asokan@acm.org}
\setcounter{Maxaffil}{0}
\renewcommand\Affilfont{\itshape\small}

\maketitle

\thispagestyle{fancy}
\chead{\small{Paper accepted to the 39th IEEE International Conference on Distributed Computing Systems (ICDCS 2019)}}
\cfoot{\small{\textcopyright 2019 IEEE. Personal use of this material is permitted. Permission from IEEE must be obtained for all other uses, in any current or
future media, including reprinting/republishing this material for advertising or promotional purposes, creating new collective works,
for resale or redistribution to servers or lists, or reuse of any copyrighted component of this work in other works.}}

\begin{abstract}
IoT devices are increasingly deployed in daily life. Many of these devices are, however, vulnerable due
to insecure design, implementation, and configuration. As a result, many
networks \textit{already} have vulnerable IoT devices that
are easy to compromise. This has led to a new category of malware
specifically targeting IoT devices. However, existing intrusion detection
techniques are not effective in detecting compromised IoT devices
given the massive scale of the problem in terms of the number of
different types of devices and manufacturers involved.

In this paper, we present \ourname, an autonomous self-learning distributed system for detecting compromised IoT devices. \ourname builds effectively on device-type-specific communication profiles \revision{without human intervention nor labeled data} that are subsequently used to detect anomalous deviations in devices' communication behavior, potentially caused by malicious adversaries. \ourname utilizes a federated learning approach for aggregating behavior profiles efficiently. To the best of our knowledge, it is the first system to employ a federated learning approach to anomaly-detection-based intrusion detection.
Consequently, \ourname can cope with emerging new and unknown attacks. We systematically and extensively evaluated more than 30 off-the-shelf IoT devices over a long term and show that \ourname is highly effective
(\SI{95.6}{\percent} detection rate) and fast ($\approx$ \SI{257}{\milli\second}) 
at detecting devices compromised by, for instance, the infamous Mirai malware. 
\ourname reported \textit{no false alarms} when evaluated in a real-world smart home deployment setting.

\end{abstract}

\section{Introduction}
\label{sect:intro}
Many new device manufacturers are entering the IoT device market, bringing out products at an ever-increasing pace. 

This ``rush-to-market'' mentality of some manufacturers has led to poor product design practices in which security considerations often remain merely an afterthought.

Consequently, many devices are released with inherent security vulnerabilities that can be exploited, which has led to an entirely new category of malware explicitly targeting IoT devices~\cite{AntonakakisUsenix2017,Edwards2016,Kolias2017,Radware2017}.

The preferred way to cope with vulnerabilities are security patches for the affected devices~\cite{Hadar2017patching}. However, many devices lack appropriate facilities for automated updates or there may be significant delays until device manufacturers provide them, mandating the use of reactive security measures like intrusion detection systems (IDS) for detecting possible device compromise ~\cite{Doshi2018DDoSDetection,krugel2002service,portnoy2001intrusion,rajasegarar2014hyperspherical}.

Signature-based IDSs look for specific communication patterns, so-called \emph{attack signatures}, associated with known attacks. Such systems are, however, unable to detect novel attacks
until the IDS vendor releases attack signatures for them~\cite{Doshi2018DDoSDetection}. 

To detect previously unknown attacks \emph{anomaly detection} needs to be used which works by profiling the normal behavior of devices and detecting attacks as deviations from this normal behavior profile~\cite{krugel2002service,portnoy2001intrusion,rajasegarar2014hyperspherical}. However, this approach often suffers from a high false alarm rate, making it unusable in practice. 
This problem is exacerbated in the IoT setting: First, there are hundreds of very heterogeneous devices on the market, which making it challenging to train a precise detection model covering all behaviors of various IoT devices.

Second, IoT devices do

typically not (notwithstanding a few exceptions) generate a lot of network traffic, as their communications are limited to, e.g., status updates about sensor readings or (relatively) infrequent interactions related to user commands. This scarcity of communications makes it challenging to train comprehensive models that can accurately cover the full behavior of IoT devices.

An effective anomaly detection model needs to capture \emph{all} benign patterns of behavior to be able to differentiate them from malicious actions. The ever-increasing number of literally thousands of types of IoT devices (ranging from temperature sensors and smart light bulbs to big appliances like  washing machines) and the typical scarcity of their communications, makes an all-encompassing behavior model 1) tedious to learn and update, and 2) too broad to be effective at detecting subtle anomalies without generating many false alarms.

\noindent\textbf{Goals and Contributions.}
To tackle the above challenges 

we present \ourname, a system for detecting compromised IoT devices.

It uses

 a novel \emph{device-type-specific} anomaly detection approach to achieve accurate detection of attacks while generating almost no false alarms. 
Major IoT device vendors, including Cisco, assisted us formulating real-world settings for our solution and usage scenarios.

We make the following contributions:
\begin{itemize}[noitemsep,topsep=0pt]
	\item \ourname, a self-learning distributed system for security monitoring of IoT devices (Sect.~\ref{sect:system-description}) based on \emph{device-type-specific detection models} for detecting anomalous device behavior:
\begin{itemize}
	
	\item It utilizes a novel anomaly detection approach based on representing network packets as symbols in a language allowing to use a language analysis technique to effectively detect anomalies (Sect.~\ref{sect:anomaly_detection}).
	\item It is the first system to apply a \emph{federated} learning approach for aggregating anomaly-detection profiles for  intrusion detection (Sect.~\ref{sect:federated-learning}). 

\end{itemize}

\item Systematic and extensive experimental analysis using more than 30 off-the-shelf IoT devices, showing that \ourname is fast (detection in $ \approx $\SI{257}{\milli\second}) and effective (\SI{95.6}{\percent} true positive rate, \emph{zero false alarms}, \revision{i.e., \SI{0}{\percent} false positive rate}) (Sect.~\ref{sect:eval-intrusion-detection}).
	\item An \Attack dataset of network traffic generated by real off-the-shelf consumer IoT devices infected with real IoT malware (Mirai~\cite{AntonakakisUsenix2017}) using which we evaluate the effectiveness of \ourname (Sect.~\ref{sect:dataset}).

\end{itemize}

We will make our datasets as well as the \ourname implementation available for research use.

\section{System Model}
\label{sect:system-description}
\begin{figure}
\centering
	\includegraphics[width=\columnwidth,trim=0 0 200 0,clip]{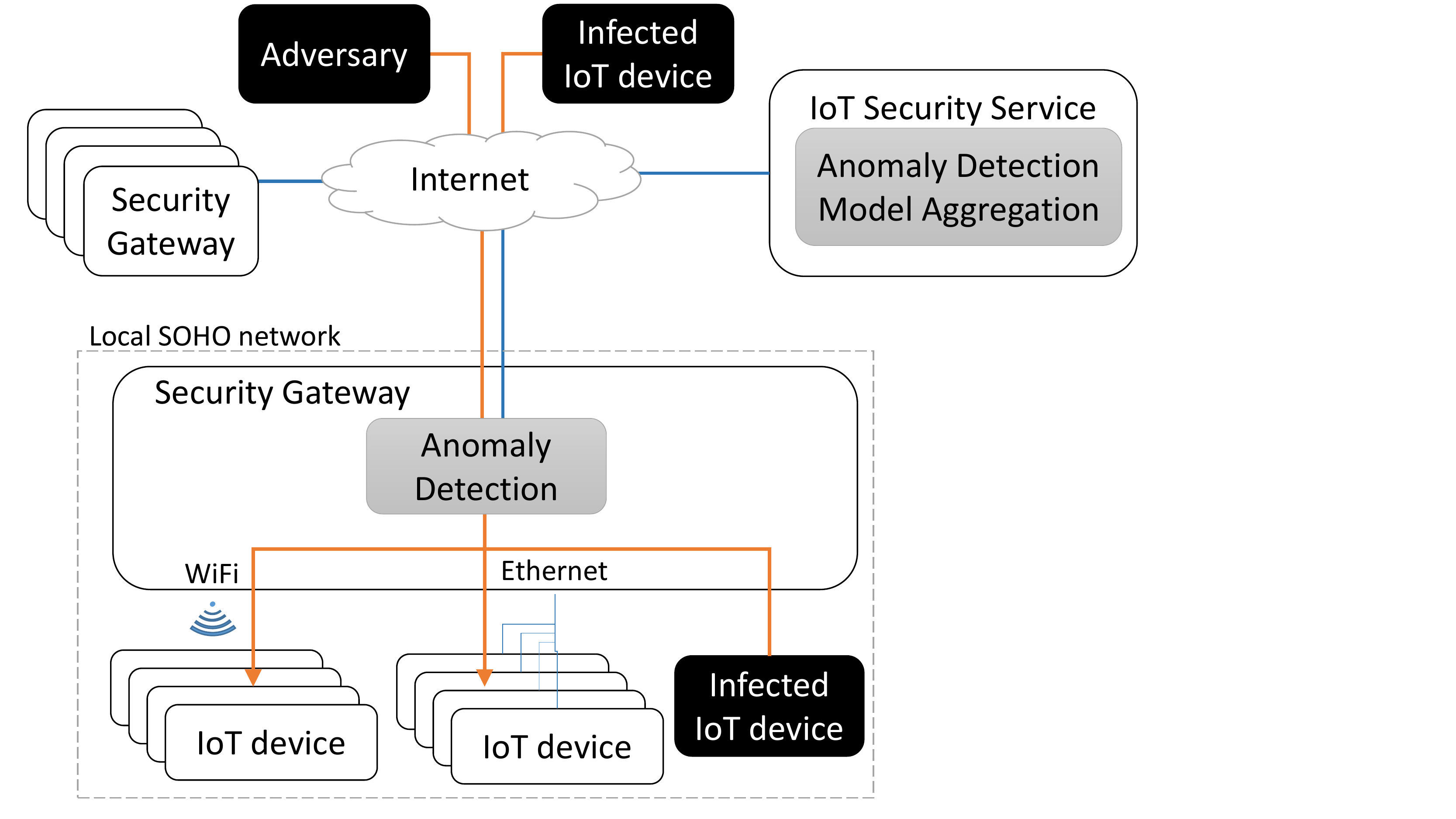}
	\caption{\ourname system model}
	\label{fig:system-model}

\end{figure}

Our system model is shown in Fig.~\ref{fig:system-model}. We consider a typical SOHO (Small Office/Home Office) network, where IoT devices connect to the Internet via an access gateway. 

\subsection{System Architecture}
The \ourname system consists of \sgw and \iotss. The role of \sgw is to monitor devices and perform anomaly detection in order to identify compromised devices in the network. It is supported by \iotss, which can be, e.g., a service provider like Microsoft, Amazon or Google that aggregates device-type-specific anomaly detection models trained by all {\sgw}s in the system.

\subsubsection{\sgw} acts as the local access gateway to the Internet to which IoT devices connect over WiFi or Ethernet. 
It hosts the \anomalydetection component. When a new device is added to the network, \sgw identifies its type as outlined in Sect.~\ref{sec:device-type-identification}. 
The \anomalydetection component monitors the communications of identified IoT devices and detects abnormal communication behavior that is potentially caused by malware (Sect.~\ref{sect:anomaly_detection}) based on anomaly detection models it trains locally and which are aggregated by the \iotss to a global detection model.

\subsubsection{\iotss} supports \sgw by maintaining a repository of device-type-specific \emph{anomaly detection models}. 
When a new device is added to the local network, \sgw identifies its \emph{device type} and retrieves the corresponding anomaly detection model for this type from \iotss. \iotss also aggregates updates to device-type-specific anomaly detection models provided by the {\sgw}s in the system.

\subsection{Adversary Model and Assumptions}
\label{sect:adversary-model}

\textbf{Adversary.} The adversary is IoT malware performing attacks against, or launching attacks from, vulnerable devices in the SOHO network. Hereby we consider all actions undertaken by the malware that it performs to discover, infect and exploit vulnerable devices as discussed in detail in Sect.~\ref{sect:attack-dataset}.

\textbf{Defense goals.} The primary goal of \ourname is to detect attacks on IoT devices in order to take appropriate countermeasures, e.g., by preventing targeted devices from being compromised or isolating compromised devices from the rest of the network. We aim to detect attacks at the earliest stage possible, preferably even \emph{before} a device can be successfully infected.
	
In addition, we make following assumptions:
\begin{itemize}[noitemsep,topsep=0pt]
	\item \textbf{\textsf{A1} - No malicious manufacturers.} IoT devices may be vulnerable but are not compromised when first released by a manufacturer.
	Adversaries must first find a vulnerability and a way to exploit it, which takes some time during which non-compromised devices generate only legitimate communications, leaving sufficient time (cf. Sect.~\ref{sect:data-needed-for-training}) to learn benign models of device behavior.

	\item \textbf{\textsf{A2} - Security Gateway is not compromised.} Since \sgw is the device enforcing security in the SOHO network, we assume that it is not compromised. Like firewall devices or antivirus software, if \sgw is compromised the SOHO network stops being protected by it. Several approaches can be used to protect it. For instance, if \sgw supports a suitable trusted execution environment, like Intel SGX~\cite{costan2016intel} or trusted platform module, its integrity can be remotely verified using \emph{remote attestation} techniques~\cite{dessouky2017fat}. 
	
	\item \textbf{\textsf{A3} - Automated identification of IoT devices.} A technique for automatically identifying and labeling IoT devices in the local SOHO network must be available. This technique should be implementable in the \sgw to identify IoT devices connected to it.
\end{itemize} 

\subsection{Device-Type Identification}
\label{sec:device-type-identification}
As \ourname uses device-type specific anomaly detection models, it requires the possibility to identify the \emph{type of devices} in the network. 
Several solutions have been designed to automatically identify and label unknown IoT devices in a network~\cite{feng2018acquisitional,Miettinen2017,audi-JSAC2019}. Alternatively, manufacturer-provided explicit device-type specifications like MUD~\cite{Lear2018} or manual labeling of IoT devices could be used.
%
%
For \ourname, \revision{we selected an existing approach - AuDI~\cite{audi-JSAC2019} that autonomously identifies the \emph{type} of individual IoT devices in a local network. 
This approach is accurate and fast (requiring only 30 minutes to identify device type at the accuracy of \SI{98.2}{\percent})}. This approach considers \emph{abstract device types} representing families of similar devices from the same device manufacturer with similar hardware and software configurations, resulting in highly identical communication behavior. 
It can be trained \emph{without the need to manually label} communication traces of pre-defined real-world device types since it works by clustering device fingerprints so that each cluster can be automatically labeled with an abstract label, e.g., $\devtype$ which represents a specific IoT device type. \revision{It justifies our assumption \textsf{A3} as mentioned above.}
 
Using this approach we can reliably map devices to a corresponding \emph{device type} for which \ourname can build a device-type-specific model of normal behavior that can be used to effectively detect anomalous deviations.
This allows \ourname to be trained and operated autonomously, \emph{without the need for human intervention at any stage}. 



\subsection{Challenges}
\label{sect:challenge}

Anomaly detection techniques 
face challenges in the IoT application scenario: 

\begin{itemize}[noitemsep,topsep=0pt]
	\item \textbf{\ChallengeDynamic - Dynamic threat landscape.} New IoT devices are released on a daily basis. A significant fraction of them have security vulnerabilities. Exploits targeting vulnerable devices are also being developed by adversaries at a similarly high pace. This makes the threats against IoT devices highly dynamic and ever-increasing.
	\item \textbf{\ChallengeResourceLimit - Resource limitations.} IoT devices have limited capabilities w.r.t. available memory, computing resources and energy often making it infeasible to perform on-device detection. 
	\item \textbf{\ChallengeHeterogeneity - IoT device heterogeneity and false alarms.} Behaviors of different IoT devices are very heterogeneous, so that anomaly detection techniques easily raise false alarms. However, to be useful in practice, anomaly detection systems must minimize false alarms.
	\item \textbf{\ChallengeScarcity - Scarcity of communications.} In contrast to high-end devices, IoT devices generate only little traffic, often triggered by infrequent user interactions. 
\end{itemize} 


\subsection{Design Choices}

\subsubsection{Gateway monitoring}

As on-device monitoring is rarely feasible due to challenge \ChallengeResourceLimit, we perform monitoring of IoT device communications on \sgw . 

\subsubsection{Device-type-specific anomaly detection} 
Since different IoT devices can have very heterogeneous behaviors (challenge \ChallengeHeterogeneity), we model each device type's behavior with a dedicated model. Consequently, each model needs to cover only the behavior of a specific device type. As IoT devices are typically single-use appliances with only a few different functions, their behavior patterns are relatively static and limited, allowing the model to accurately capture all possible legitimate behaviors of a device type. Thus, the model 
is less prone to trigger false alarms (details in Sect.~\ref{sec:device-typ-specific}), thereby effectively addressing challenge \ChallengeHeterogeneity.

\subsubsection{Federated learning approach}
Anomaly detection models are learned using a \emph{federated learning approach} where {\sgw}s use locally collected data to train local models which \iotss aggregates into a global model (details in Sect.~\ref{sect:federated-learning}). This aggregation maximizes the usage of limited information obtained from scarce communications at each gateway (challenge \ChallengeScarcity) and helps to improve the accuracy of anomaly detection models by utilizing all available data for learning.

\subsubsection{Autonomous self-learning}
\label{sect:bg-deviceID}
Anomaly detection models are trained using data autonomously labeled with the device-type that generated it. Device types are also learned and assigned in an autonomous manner.
The whole process does therefore not require any human intervention, which allows \ourname to respond quickly and autonomously to new threats, addressing challenge \ChallengeDynamic. It is worth noting that \ourname starts operating with no anomaly detection model. It learns and improves these models as {\sgw}s aggregate more data.



\subsubsection{Modeling techniques requiring little data} As discussed in detail in Sect.~\ref{sect:anomaly_detection}, we select features and machine learning algorithms (GRU) that can be efficiently trained even with few training data. This design choice addresses challenge \ChallengeScarcity.

%

\section{Device-Type-Specific Anomaly Detection}
\label{sect:anomaly_detection}

\begin{figure}
		\centering
	\includegraphics[width=0.65\columnwidth]{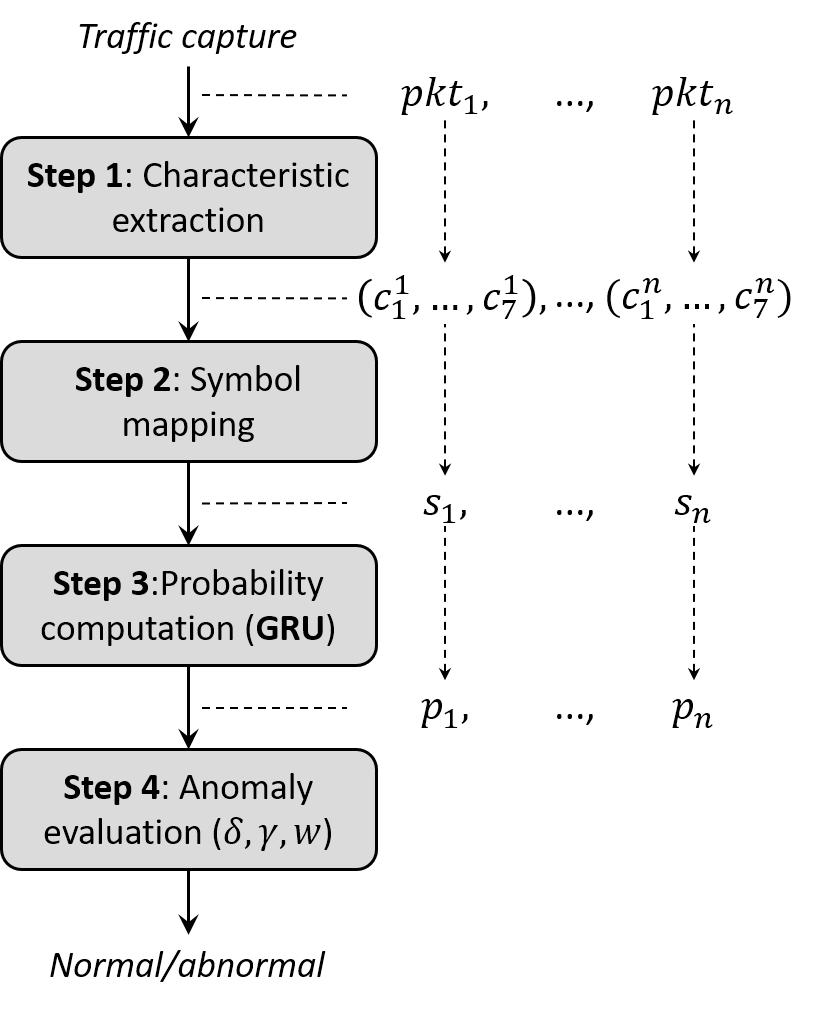}
	\caption{Overview of device-type-specific anomaly detection}
	\label{fig:anomaly-detection-flowchart}
\end{figure}

Our anomaly detection approach is based on evaluating the communication patterns of a device to determine whether it is consistent with the learned benign communication patterns of that particular device type. The detection process is shown in Fig.~\ref{fig:anomaly-detection-flowchart}. In \textbf{Step 1} the communication between the \sgw and the IoT device is captured as a sequence of packets $\pkt_1, \pkt_2, \ldots $. Each packet $ \pkt_i $ is then in \textbf{Step 2} mapped to a corresponding symbol $ \sym_i $ characterizing the type of the packet using a mapping that is based on distinct characteristics $ c_1, c_2, \ldots, c_7 $ derived from each packet's header information as discussed in Sect.~\ref{sect:packet-flow-modelling}. The mapped sequence of symbols $ \sym_1, \sym_2, \ldots $ is then in \textbf{Step 3} input into a pre-trained model using Gated Recurrent Units (GRUs)~\cite{GRU_libs2014,Chung_GRU2014}. The GRU model will calculate a probability estimate $ p_i $ for each symbol $ s_i $ based on the sequence of $ k $ preceding symbols $ s_{i-k}, s_{i-k+1}, \ldots, s_{i-1} $. GRU is a novel approach to recurrent neural networks (RNN) currently being a target of lively research. GRUs provide similar accuracy as other RNN approaches but are computationally less expensive~\cite{Chung_GRU2014, Wu2017RRN}.
In \textbf{Step 4} the sequence of occurrence probability estimates $ \prob_1, \prob_2, \ldots $ is evaluated to determine possible anomalies. If the occurrence probabilities $\prob_i$ of a sufficient number of packets in a window of consecutive packets fall below a detection threshold, as described in detail in Sect.~\ref{sect:detection}, the packet sequence is deemed anomalous and an alarm is raised.

\subsection{Modelling Packet Sequences}
\label{sect:packet-flow-modelling}
Data packets $\pkt_i$ in the packet sequence $ \pkt_1, \pkt_2, \ldots $ emitted by an IoT device are mapped into \emph{packet symbols} $\sym_i $ based on 7-tuples $ ({c_1},{c_2},\ldots,{c_7})$ of discrete packet characteristics $ c_i $ of packet $ \pkt_i $.
This mapping is defined by a device-type-specific mapping function $ \map: A \rightarrow B_\devtype$ s.t. $ \map(\pkt_i) = \sym_i$ where $A$ is the domain of raw network packets $\pkt$ and $B_\devtype$ is the domain of packet symbols $\sym$ for device-type $\devtype$. 
Mapping $ \map $ assigns each unique combination of packet characteristics $ (c_1, \ldots, c_7) $ a dedicated symbol $ \sym $ representing the 'type' of the particular packet.
We use the following packet characteristics shown also in Tab.~\ref{tab:symbol-mapping}:
\begin{itemize}[noitemsep]
	\item \textbf{$c_1$ direction:} (\emph{incoming / outgoing}) Normal TCP traffic is usually balanced two-way communication but abnormal is not as, e.g., a bot only sends packets to a victim without receiving replies when running DDoS attacks.
	\item \textbf{$c_2$ and $c_3$ local and remote port type:} (\emph{system / user / dynamic}) Each device-type uses specific ports designed by the manufacturers while malicious attack patterns usually use different ports.
	\item \textbf{$c_4$ packet length:} (\emph{bin index of packet's length where eight most frequently occurring packet lengths receive dedicated bins and one bin for other packet length values}) Each device-type communicates using specific packet patterns with specific packet lengths that are mostly different in malicious attack patterns.
	\item \textbf{$c_5$ TCP flags:} Normal communications contain packets with specific TCP flag sequences e.g., $SYN \rightarrow SYNACK \rightarrow ACK \rightarrow PUSH \rightarrow FIN $. However, many attacks do not follow standard protocols, e.g., SYN flood (DDoS attack) only sends $ SYN $ messages.
	\item \textbf{$c_6$ encapsulated protocol types:} Each device type usually uses a set of specific protocols, which is likely different from protocol types used in attacks.
	\item \textbf{$c_7$ IAT bin:} (\emph{bin index of packet inter-arrival time (IAT) using three bins: $<$ \SI{0.001}{\milli\second}, \SIrange{0.001}{0.05}{\milli\second}, and $>$ \SI{0.05}{\milli\second}}) Many attacks (e.g., DDoS) usually generate traffic at a high packet rate, resulting in smaller IAT values in than normal communications.
\end{itemize}

\begin{table}[htb]
	\caption{Packet characteristics used in symbol mapping}
	\label{tab:symbol-mapping}
	\begin{tabular}{cll}
		\hline 
		ID & Characteristic & Value \\ 
		\hline
		$c_1$ & direction & $ 1= $ incoming, $ 0= $ outgoing \\ 
		$c_2$ & local port type & bin index of port type \\ 
		$c_3$ & remote port type & bin index of port type \\ 
		$c_4$ & packet length & bin index of packet length \\ 
		$c_5$ & TCP flags & TCP flag values\\ 
		$c_6$ & protocols & encapsulated protocol types \\ 
		$c_7$ & IAT bin & bin index of packet inter-arrival time \\
		\hline 
	\end{tabular} 
\end{table}



\subsection{Detection Process}
\label{sect:detection}
\begin{figure}
	\centering
	\subfloat[Symbol distribution]{\includegraphics[width=\columnwidth]{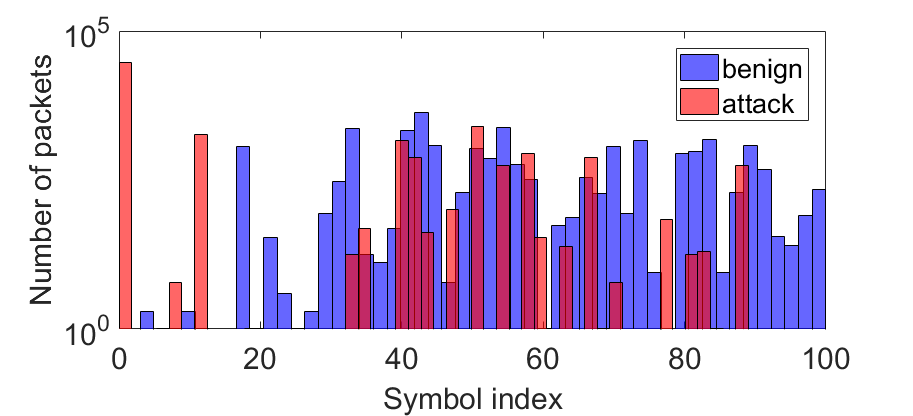}\label{fig:symbol_distribution}}
	\\
	\subfloat[Probability distribution ]{\includegraphics[width=\columnwidth]{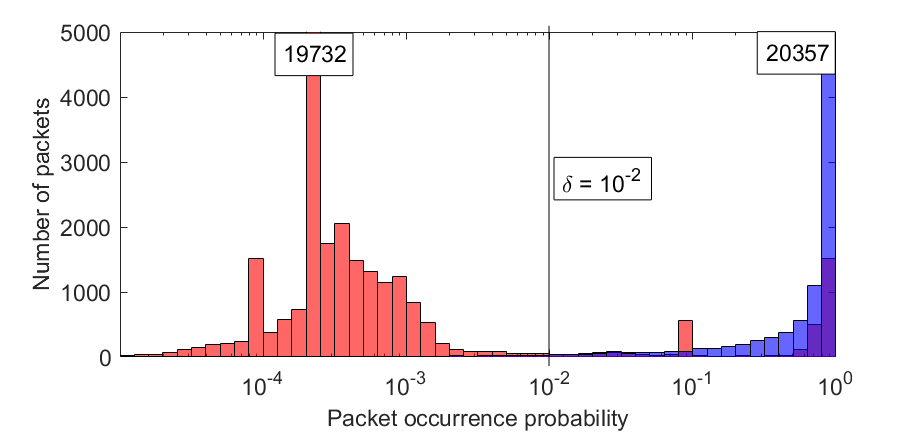}\label{fig:probability_distribution}}
	\caption{Packet symbol occurrence frequencies and occurrence probability estimates for benign and attack traffic for Edimax smart power plugs}
	\label{fig:symbol-occurrence-distribution}
\end{figure}

Fig.~\ref{fig:symbol_distribution} shows an example of the occurrence frequencies of individual packet symbols for benign and attack traffic (as generated by the Mirai malware) for Edimax smart power plugs. It can be seen that using packet symbols alone to distinguish between benign and attack traffic is not sufficient, as both traffic types contain packet types that are mapped to the same symbols. Our detection approach is therefore based on estimating the likelihood of observing individual packet types given the sequence of preceding packets. The rationale behind this approach is the observation that IoT device communications usually follow particular characteristic patterns. Traffic generated by IoT malware, however, doesn't follow these patterns and can therefore be detected.
We will thus use the detection model to calculate an \emph{occurrence probability} $\prob_i$ for each packet symbol $s_i$ given the sequence of $k$ preceding symbols $< s_{i-k}, s_{i-k+1}, \ldots, s_{i-1} >$, i.e.,
\begin{equation}
\label{eq:probability}
\prob_i = P(s_i\vert < s_{i-k}, s_{i-k+1}, \ldots, s_{i-1} >) 
\end{equation}
Parameter $k$ is a property of the used GRU network and denotes the length of the lookback history, i.e., the number of preceding symbols that the GRU takes into account when calculating the probability estimate.
From Fig.~\ref{fig:probability_distribution} we can see that these probability estimates are on average higher for packets belonging to benign traffic patterns, and lower for packets generated by malware on an infected device and can therefore be flagged as anomalous.
%
%
%

\begin{figure}
	\includegraphics[width=\columnwidth]{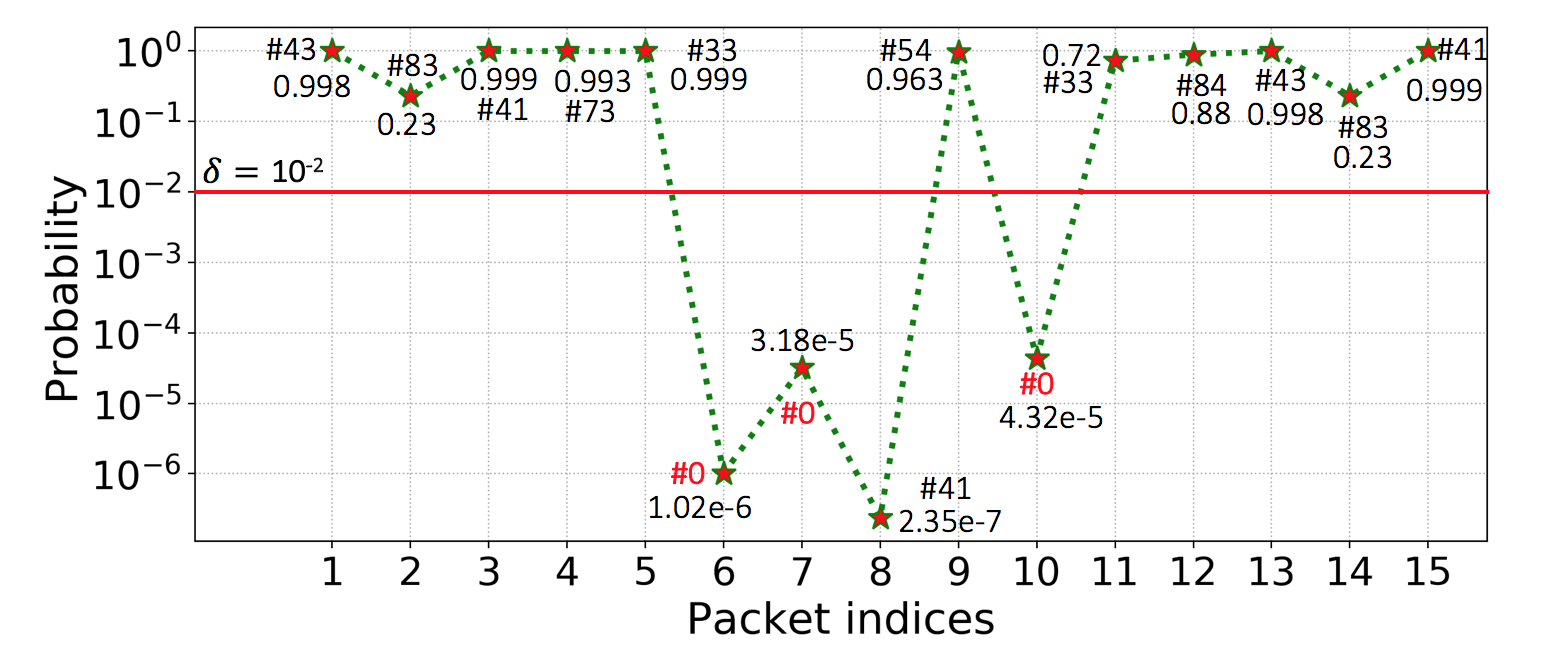}
	\caption{Occurrence probabilities of 15 packets from Edimax Plug when Mirai was in standby stage. The red '\#0' denotes the malicious packets.}
	\label{fig:OccuranceProb}
\end{figure}

\begin{definition}[Anomalous packets]
	\label{def:packet-anomaly}
	Packet $\pkt_i$ mapped to packet symbol $\sym_i$ is \emph{anomalous}, if its occurrence probability $\prob_i$ is below \emph{detection threshold} $\detThr$, i.e., if 
	\begin{equation}
	\label{eq:packet-seq-anomaly}
	\prob_i < \detThr
	\end{equation}
\end{definition}
We performed an extensive empirical analysis of the probability estimates provided by device-specific detection models for both benign and attack traffic for the datasets described in Sect.~\ref{sect:eval-intrusion-detection} and could determine that a value of $\detThr = 10^{-2}$ provides a good separation between benign and attack traffic, as can be also seen in Fig.~\ref{fig:probability_distribution}.
An example of our approach is shown in Fig.~\ref{fig:OccuranceProb}. Malicious packets (represented by symbol '\#0') get very low probability estimates (\textless $10^{-4}$), distinguishing them clearly from benign packets. However, their presence at indices $ 6-7 $ also affects the estimate of the benign packet '\#41' at index $ 8 $ (\textless $10^{-6}$), since the sequence of packets preceding this packet is unknown to the detection model.

Triggering an anomaly each time an anomalous packet is observed would lead to numerous false positive detections, as also benign traffic may contain noise that is not covered by the GRU model and will therefore receive low occurrence probability estimates. An anomaly is therefore triggered only in the case that a significant number of packets in a window of consecutive packets are anomalous.

\begin{definition}[Anomaly triggering condition]
	\label{def:anomaly-trigger-condition}
	Given a window $W$ of $\win$ consecutive packets $W = (\pkt_1, \pkt_2, \ldots, \pkt_w)$ represented by symbol sequence $S = (\sym_1, \sym_2, \ldots \sym_\win)$, we trigger an anomaly alarm, if the fraction of anomalous packets in $W$ is larger than an anomaly triggering threshold $\anomalyThr$, i.e., if
	\begin{equation}
	\label{eq:anomaly-trigger-condition}
	\frac{\vert\{\sym_i \in S \vert \prob_i < \detThr \}\vert}{\win} > \anomalyThr
	\end{equation}
\end{definition}

\section{Federated Learning Approach}
\label{sect:federated-learning}
The GRU models are learned using traffic collected at several {\sgw}s, each monitoring a client IoT network. Each \sgw observing a device of a particular type $\devtype$ contributes to training its anomaly detection model. We take a \textit{federated learning} approach to implement the distributed learning of models from several clients. 
Federated learning is a communication-efficient and privacy-preserving learning approach suited for distributed optimization of Deep Neural Networks (DNN) ~\cite{KonecnyMYRSB16,NIPS2017_7029}. In federated learning, clients do not share their training data but rather train a local model and send model updates to a centralized entity which aggregates them.
Federated learning is chosen because it is suitable~\cite{mcmahan2017communication} for scenarios where:
\begin{itemize}[noitemsep]
	\item data are massively distributed, so that there is a large number of clients each having a small amount of data. IoT devices typically generate little traffic, which means only little data can be provided by each client alone.
	\item contributions from clients are imbalanced. In our system, the training data available at each \sgw depends on the duration that an IoT device has been in the network and the amount of interaction it has had, which varies largely between clients.
\end{itemize} 

\subsection{Learning Process}
\label{sect:federated-learning-process}
\begin{figure}
	\centering
	\includegraphics[width=0.9\columnwidth]{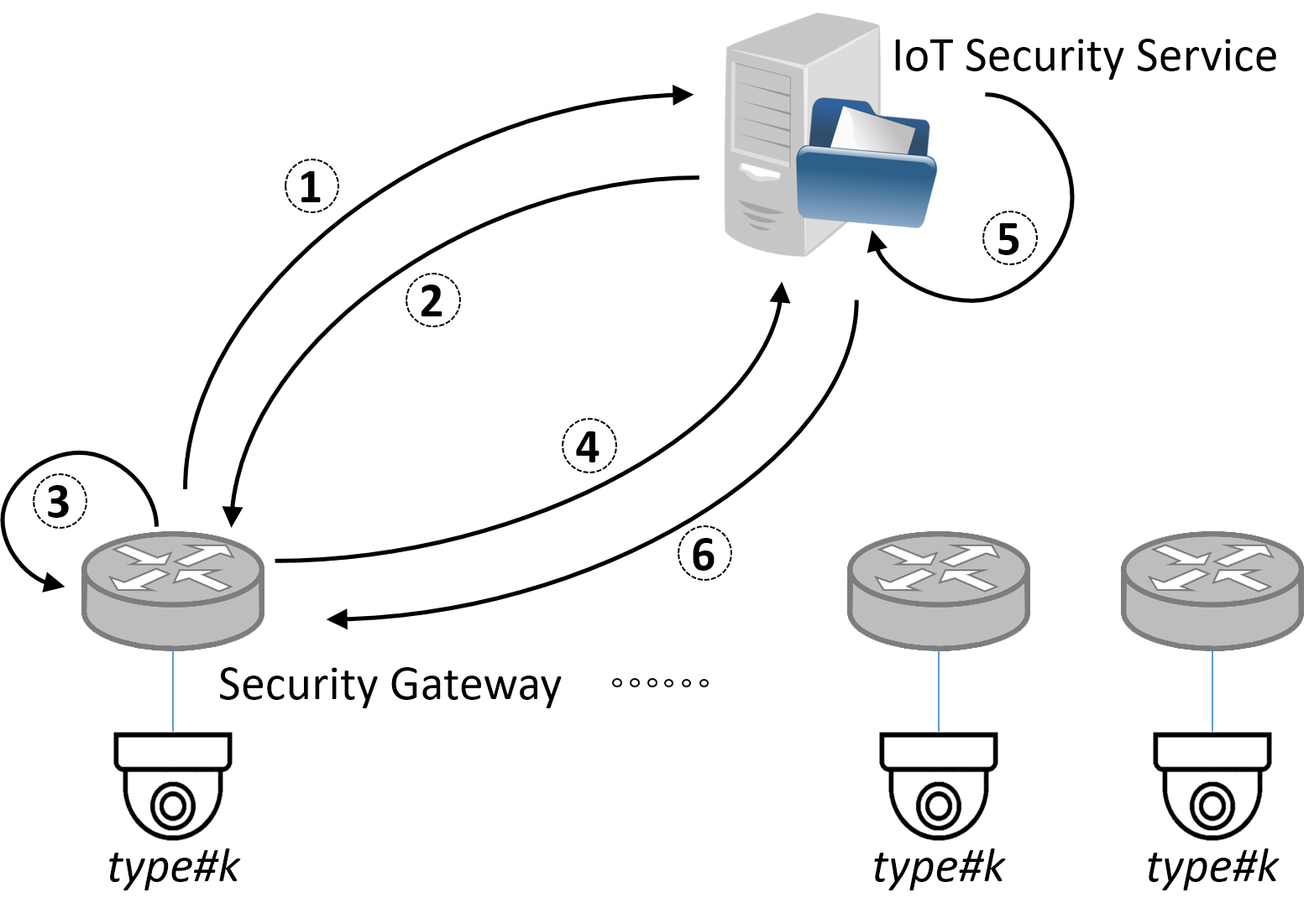}
	\caption{Overview of federated learning process}
	\label{fig:federated-flowchart}
\end{figure}
The federated training process is illustrated in Fig.~\ref{fig:federated-flowchart}. Each \sgw having devices of a particular type $ \devtype $ in its network requests a detection profile for this type from \iotss in \textbf{Step 1} and gets an initial GRU model for this type in \textbf{Step 2}. At the start of \ourname, this model is random, otherwise it is already trained through several rounds of the following process.
In \textbf{Step 3} the global model is re-trained locally by each \sgw with traces collected by monitoring communication of the $\devtype$ devices. 
Then in \textbf{Step 4} local updates made to the model by each \sgw are reported to \iotss which in \textbf{Step 5} aggregates them as defined in Def.\ref{def:model-aggregation} ~\cite{mcmahan2017communication} to improve the global model.
Finally, the updated global model for $\devtype$ devices is then pushed back to \sgw and used for anomaly detection (\textbf{Step 6}).
The re-training of the model is performed on a regular basis to improve its accuracy. 

\revision{
\begin{definition}[Global Model Aggregation]
	\label{def:model-aggregation}
	Given n clients with their associated model weights $W_1, \ldots, W_n$ trained by associated number of data samples $s_1, \ldots, s_n$. We define the global model $G$ which is aggregated from those local models as follows:
	\begin{equation}
	\label{eq:model-aggregation}
	G = \sum_{i=1}^{n}\frac{s_i}{s}{W_i} \quad ( \textrm{where} \; s = \sum_{i=1}^{n}{s_i}) 		
	\end{equation}
\end{definition}
}

To train our models we adopt an approach introduced by McMahan \emph{et al.}~\cite{mcmahan2017communication}. Each client (\sgw) trains its GRU model locally for several epochs before reporting updates to \iotss. This limits the communication overhead by reducing the number of updates to send to the \iotss. 
To the best of our knowledge we are the first to employ a federated learning approach for anomaly detection-based intrusion detection. 
\subsection{Federated Learning Setup}
\label{sect:federated-learning-setup}
We implemented the federated learning algorithm utilizing the \emph{flask}~\cite{flask} and \emph{flask\_socketio}~\cite{flask-sio} libraries for the server-side application and the \emph{socketIO-client}~\cite{sio-client} library for the client-side application.
The socketIO-client uses the \emph{gevent} asynchronous framework~\cite{gevent} which provides a clean API for concurrency and network related tasks. We used the \emph{Keras}~\cite{keras} library with \emph{Tensorflow} backend to implement the GRU network with the parameters selected in Sect.~\ref{sect:parameter-selection}.
 
 \section{Experimental Setup}
\label{sect:experimental-setup}
To evaluate \ourname, we apply it on the use case of detecting real-life IoT malware. We selected Mirai for this purpose, since its source code is publicly available and several infamous malware variants like Hajime~\cite{Edwards2016} or Persirai~\cite{Yeh2017} have been implemented using the same code base. Mirai also realizes similar attack stages (detailed in Sect.~\ref{sect:attack-dataset} below) as state-of-the-art IoT malware~\cite{pa2016iotpot,Kolias2017}. This makes Mirai a highly relevant baseline for IoT malware behavior.

\subsection{Datasets}
\label{sect:dataset}

\begin{table*}
	\centering
	\caption{33 IoT devices used in the \Activity, \Deployment and \Attack datasets and their connectivity technologies + Affectation of these devices to 23 \ourname device types during evaluation.}
	\label{tab:aggreg-type}
	\begin{tabular}[th]{lllccc|ccc}
			\textbf{Device-type} & \textbf{Identifier} & \textbf{Device model} & \mcrot{1}{l}{60}{WiFi} & \mcrot{1}{l}{60}{Ethernet} & \mcrot{1}{l}{60}{Other} & \mcrot{1}{l}{60}{\Activity} & \mcrot{1}{l}{60}{\Deployment} & \mcrot{1}{l}{60}{\Attack}\\
		\hline
		\textit{type\#01} & ApexisCam & Apexis IP Camera APM-J011 & \bul & \bul & \cir & \bul & \cir & \cir \\
		\textit{type\#02} &CamHi & Cooau Megapixel IP Camera & \bul & \bul & \cir & \bul & \cir & \cir \\
		\textit{type\#03}    & D-LinkCamDCH935L & D-Link HD IP Camera DCH-935L & \bul & \cir & \cir & \bul & \cir & \cir \\ \hline
		 \multirow{2}{*}{\textit{type\#04} }  	& D-LinkCamDCS930L & D-Link WiFi Day Camera DCS-930L & \bul & \bul & \cir & \bul  & \bul & \bul  \\
				& D-LinkCamDCS932L & D-Link WiFi Camera DCS-932L & \bul & \bul & \cir & \bul  & \bul & \bul  \\ \hline
				& D-LinkDoorSensor & D-Link Door \& Window sensor & \cir & \cir & \bul & \bul & \cir & \cir  \\
				& D-LinkSensor & D-Link WiFi Motion sensor DCH-S150 & \bul & \cir & \cir & \bul & \bul  & \cir \\
		\textit{type\#05}  	& D-LinkSiren	& D-Link Siren DCH-S220 & \bul & \cir & \cir & \bul & \cir  & \cir \\
				& D-LinkSwitch & D-Link Smart plug DSP-W215 & \bul & \cir & \cir & \bul & \bul  & \cir \\
				& D-LinkWaterSensor & D-Link Water sensor DCH-S160 & \bul & \cir & \cir & \bul & \cir & \cir \\ \hline
		\multirow{2}{*}{\textit{type\#06} } 	& EdimaxCamIC3115 & Edimax IC-3115W Smart HD WiFi Network Camera  & \bul & \bul & \cir & \bul & \cir & \cir \\
				& EdimaxCamIC3115(2) & Edimax IC-3115W Smart HD WiFi Network Camera  & \bul & \bul & \cir & \bul & \cir & \cir  \\ \hline
		 \multirow{2}{*}{\textit{type\#07} } & EdimaxPlug1101W & Edimax SP-1101W Smart Plug Switch & \bul & \bul & \cir & \bul & \bul & \bul  \\
				& EdimaxPlug2101W & Edimax SP-2101W Smart Plug Switch & \bul & \cir & \cir & \bul & \bul & \bul \\ \hline
		\textit{type\#08}  	& EdnetCam & Ednet Wireless indoor IP camera Cube  & \bul & \bul & \cir & \bul & \cir & \cir \\
		\textit{type\#09}  	& EdnetGateway & Ednet.living Starter kit power Gateway  & \bul & \cir & \bul & \bul & \bul  & \cir \\
		\textit{type\#10}  	& HomeMaticPlug & Homematic pluggable switch HMIP-PS & \cir & \cir & \bul & \bul & \cir  & \cir \\
		\textit{type\#11}  	& Lightify & Osram Lightify Gateway & \bul & \cir & \bul & \bul & \bul & \cir \\
		\textit{type\#12}  	& SmcRouter & SMC router SMCWBR14S-N4 EU & \bul & \bul & \cir & \bul & \cir & \cir  \\   \hline
		 \multirow{2}{*}{\textit{type\#13} }  & TP-LinkPlugHS100 & TP-Link WiFi Smart plug HS100 & \bul & \cir & \cir & \bul & \cir & \cir  \\ 
				& TP-LinkPlugHS110 & TP-Link WiFi Smart plug HS110 & \bul & \cir & \cir & \bul & \cir  & \cir \\ \hline
		\textit{type\#14}  	& UbnTAirRouter & Ubnt airRouter HP & \bul & \bul & \cir & \bul & \cir  & \bul \\
		\textit{type\#15}  	& WansviewCam & Wansview 720p HD Wireless IP Camera K2 & \bul & \cir & \cir & \bul & \cir & \cir  \\
		\textit{type\#16}  	& WeMoLink & WeMo Link Lighting Bridge model F7C031vf & \bul & \cir & \bul & \bul & \cir & \cir \\ \hline
		\multirow{2}{*}{\textit{type\#17} } & WeMoInsightSwitch & WeMo Insight Switch model F7C029de & \bul & \cir & \cir & \bul & \cir & \cir   \\
				& WeMoSwitch & WeMo Switch model F7C027de& \bul & \cir & \cir & \bul & \cir  & \cir \\ \hline
		\textit{type\#18}  	& HueSwitch & Philips Hue Light Switch PTM 215Z  & \cir & \cir & \bul & \bul & \bul & \cir   \\
		\textit{type\#19}  	& AmazonEcho & Amazon Echo & \bul & \cir & \cir & \bul &  \bul  & \cir \\
		\textit{type\#20}  	& AmazonEchoDot & Amazon Echo Dot & \bul & \cir & \cir & \bul & \cir & \cir \\
		\textit{type\#21}  	& GoogleHome & Google Home & \bul & \cir & \cir & \bul & \bul & \cir   \\
		\textit{type\#22}  	& Netatmo & Netatmo weather station with wind gauge & \bul & \cir & \bul & \bul & \bul & \cir  \\ \hline
		\multirow{2}{*}{\textit{type\#23}} 	& iKettle2 & Smarter iKettle 2.0 water kettle SMK20-EU & \bul & \cir & \cir & \bul & \bul & \cir  \\
				& SmarterCoffee & Smarter SmarterCoffee coffee machine SMC10-EU & \bul & \cir & \cir & \bul & \bul  & \cir  \\ \hline
	\end{tabular}
\end{table*}

\begin{table}
	\centering      
	\caption{Characteristics of used datasets}
	\label{tab:dataset}
	\begin{tabular}{lrrrr}
		\multicolumn{1}{c}{\begin{tabular}[c]{@{}c@{}}Dataset\\ (Number of devices)\end{tabular}} & \multicolumn{1}{c}{\begin{tabular}[c]{@{}c@{}}Time\\ (hours)\end{tabular}} & \multicolumn{1}{c}{\begin{tabular}[c]{@{}c@{}}Size\\ (MiB)\end{tabular}} & \multicolumn{1}{c}{Flows} & \multicolumn{1}{c}{Packets} \\ \hline
		\Activity (33)                                                                             & 165                                                                        & 465                                                                      & 115,951                   & 2,087,280                   \\
		\Deployment (14)                                                                           & 2,352                                                                         & 578                                                                      & 95,518                    & 2,286,697                   \\
		\Attack (5)                                                                                & 84                                                                         & 7,734                                                                    & 8,464,434                 & 21,919,273                  \\ \hline
\end{tabular}
\end{table}

\begin{figure}
	\includegraphics[width=1\linewidth]{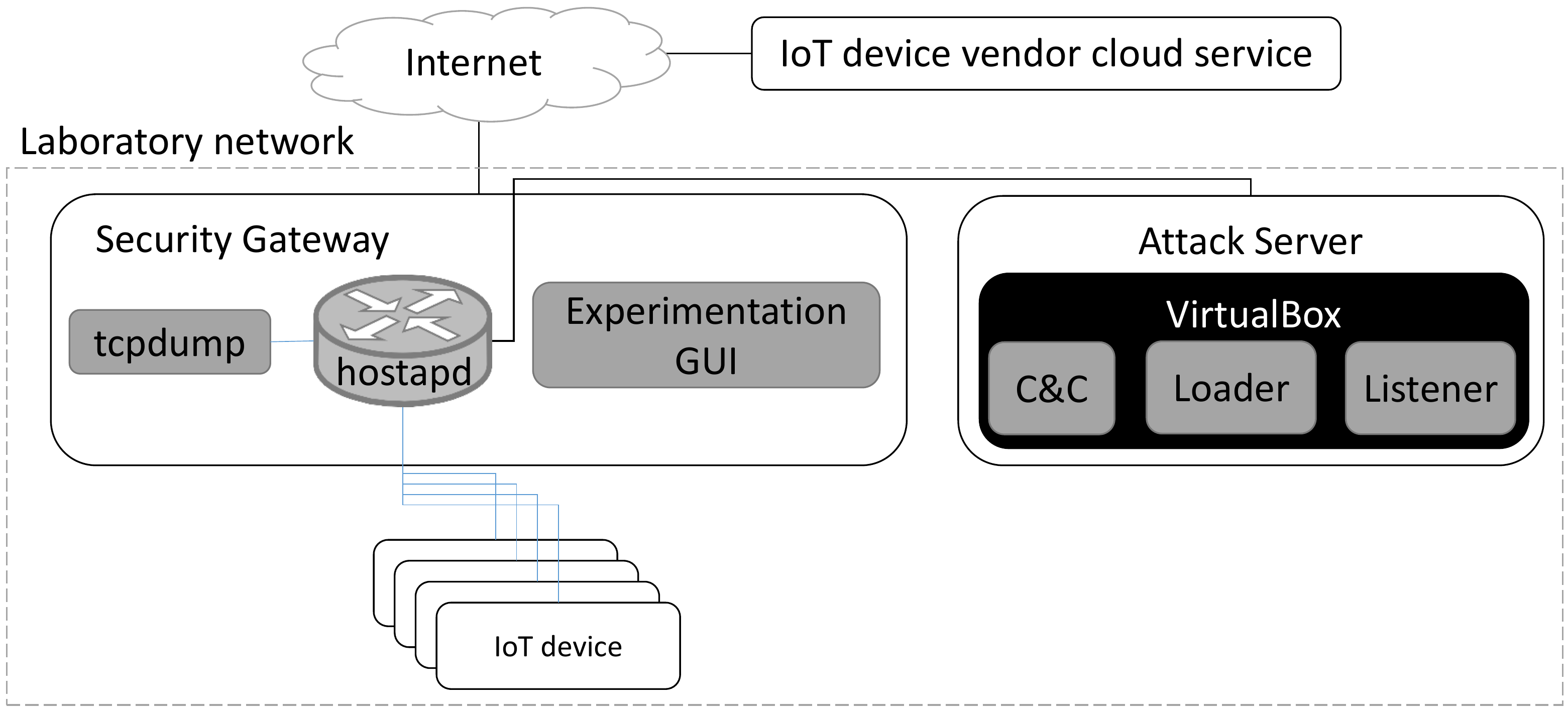}
	\caption{Laboratory network setup}
	\label{fig:lab-setup}
\end{figure}

We collected extensive datasets about the communication behavior of IoT devices in laboratory and real-world deployment settings. The monitored devices included 33 typical consumer IoT devices like IP cameras, smart power plugs and light bulbs, sensors, etc. The devices were mapped by our device-type-identification method to 23 unique device types. The detailed list of devices and assignment to device-types can be found in Tab. \ref{tab:aggreg-type}. We collected datasets by setting up a laboratory network as shown in Fig. \ref{fig:lab-setup} using \texttt{hostapd} on a laptop running Kali Linux to create a gateway acting as an access point with WiFi and Ethernet interfaces to which IoT devices were connected. On the gateway we collected all network traffic packets originating from the monitored devices using \texttt{tcpdump}.


%

\subsubsection{\Activity dataset}
A key characteristic of IoT devices is that they expose only a few distinct actions accessible to users, e.g., ON, OFF, ADJUST, etc. 
To capture the communication patterns related to user interactions with IoT devices, we collected a dataset encompassing all such actions being invoked on the respective IoT devices. We repeatedly performed actions shown in Tab.~\ref{tab:iot-actions}. Each of the actions was repeated 20 times (20-time repetition chosen as a rule of thumb). To capture also less intensive usage patterns, the dataset was augmented with longer measurements of two to three hours, during which actions were triggered only occasionally. This dataset contains data from 33 IoT devices out of which 27 have both action and standby data. Six devices (lighting and home automation hubs) have standby data only because they do not provide meaningful actions that users could invoke.

\begin{table}
	\begin{center}
		\caption{Actions for different IoT device categories}
		\label{tab:iot-actions}
		\begin{tabular}{lp{5.5cm}}
			Category (count)& Typical actions\\
			\hline
			IP cameras (6)	& START / STOP video, adjust settings, reboot\\ 
			Smart plugs	(9) & ON, OFF, meter reading \\ 
			Sensors (3)	& trigger sensing action \\ 
			Smart lights (4) & turn ON, turn OFF, adjust brightness\\
			Actuators (1) & turn ON, turn OFF\\
			Appliances (2) & turn ON, turn OFF, adjust settings\\
			Routers	(2) & browse amazon.com \\ 
			Hub devices (6) & no actions\\
			\hline
		\end{tabular} 
	\end{center}	
\end{table}

\subsubsection{\Deployment dataset}
To evaluate \ourname in a realistic smart home deployment setting, in particular with regard to how many false alarms it will raise, we installed a number of ($ n=14 $) different smart home IoT devices\footnote{The number of devices was limited, as the driver of the used WiFi interface allowed at most 16 devices to reliably connect to it simultaneously.} in several different domestic deployment scenarios. This deployment involved real users and collected communication traces of these devices under realistic usage conditions.
We used the same set-up as in the laboratory network for the domestic deployment, albeit we excluded the attack server. 
Users used and interacted with the IoT devices as part of their everyday life. Packet traces were collected continuously during one week.

\subsubsection{\Attack dataset}
\label{sect:attack-dataset}
For evaluating the effectiveness of \ourname at detecting attacks, we collected a dataset comprising malicious traffic of IoT devices infected with Mirai malware~\cite{AntonakakisUsenix2017,Kolias2017} in all four different attack stages discussed below: \emph{pre-infection}, \emph{infection}, \emph{scanning} and \emph{DoS attacks} (as a monetization stage). Additionally, we collected traffic when Mirai was in a \textit{standby} mode, i.e., not performing any attack but awaiting commands from its \emph{Command \& Control} server. 

Among 33 experimental devices, we found 5 devices which are vulnerable to the Mirai malware. The  \Attack dataset was collected from those five devices: \emph{D-LinkCamDCS930L}, \emph{D-LinkCamDCS932L}, \emph{EdimaxPlug1101W}, \emph{EdimaxPlug2101W} and \emph{UbntAirRouter}. This was done by installing the \emph{Command \& Control}, \emph{Loader} and \emph{Listener} server modules on the laboratory network for infecting target devices with Mirai and controlling them. Infection was achieved using security vulnerabilities like easy-to-guess default passwords to open a terminal session to the device and issuing appropriate commands to download the malware binary onto the device.


In the \emph{pre-infection} stage, \emph{Loader} sends a set of commands via telnet to the vulnerable IoT device to prepare its environment and identify an appropriate method for uploading the Mirai binary files.
We repeated the pre-infection process 50 times for each device. During each run, around 900 pre-infection-related packets were generated.

After pre-infection the \emph{infection stage} commences, during which \emph{Loader} uploads Mirai binary files to the IoT device. It supports three upload methods: \texttt{wget}, \texttt{tftp} and \texttt{echo} (in this priority order). To infect the two D-Link cameras and the Ubnt router \emph{Loader} uses \texttt{wget}, on the Edimax plugs it will resort to using \texttt{tftp} as these are installed on the devices by default.
We repeated the infection process 50 times for each device, each run generating approximately 700 data packets.    

In the \emph{scanning stage} we collected packets while the infected devices were actively performing a network scan in order to locate other vulnerable devices. Data collection was performed for five minutes per device, resulting in a dataset of more than 446,000 scanning data packets.

\revision{We extensively tested the \emph{DoS attack stage}, utilizing all ten different DoS attack vectors (for details, see~\cite{diot-arXiv}) available in the Mirai source code~\cite{Mirai2018}.} We ran all attacks separately on all five compromised devices for five minutes each, generating more than 20 million packets of attack traffic in total.



Tab.~\ref{tab:dataset} summarizes the sizes and numbers of distinct packets and packet flows in the different datasets. While packet flows can't be directly mapped to distinct device actions, they do provide a rough estimate of the overall level of activity of the targeted devices in the dataset. 

\subsection{Parameter Selection}
\label{sect:parameter-selection}
Based on initial experiments with our datasets (Tab.~\ref{tab:dataset}) we inferred that a lookback history of $k = 20$ symbols is sufficient to capture most communication interactions with sufficient accuracy. 
We used a GRU network with three hidden layers of size 128 neurons each. The size of the input and output layers is device-type-specific and equal to the number of mapping symbols of the function $\map$, which is equal to $\mid B_\devtype \mid$ (cf. Sect.~\ref{sect:packet-flow-modelling}). 
We learned 23 anomaly detection models, each corresponding to a device type identified using the method described in Sect.~\ref{sec:device-type-identification}. Each anomaly detection model was trained with, and respectively tested on, communication from all devices matching the considered type. 

\subsection{Evaluation Metrics}
\label{sect:metric}
We use false positive and true positive rate (FPR and TPR) as measures of fitness. FPR measures the rate at which benign communication is incorrectly classified as anomalous by our method causing a false alarm to be raised. TPR is the rate at which attacks are correctly reported as anomalous. We seek to minimize FPR, since otherwise the system easily becomes unusable, as the user would be overwhelmed with false alarms. At the same time we want to maximize TPR so that as many attacks as possible will be detected by our approach.


Testing for false positives was performed by four-fold cross-validation for device types in the \Activity and \Deployment datasets. The data were divided equally into four folds using three folds for training and one for testing. To determine the FPR, we divided the testing dataset according to Def.~\ref{def:anomaly-trigger-condition} into windows of $w=250$ packets. Since the testing data contained only benign communications, any triggered anomaly alarm for packets of the window indicated it as a false positive, whereas windows without alarms were considered a true negative.

Testing for true positives was done by using the \Activity and \Deployment datasets as training data and the \Attack dataset for testing with the same settings as for false positive testing. Moreover, as we know that the \Attack dataset also contains benign traffic corresponding to normal operations of the IoT devices, we were interested in the average duration until detection. Therefore, in each window of $ w=250 $ packets we calculated the number of packets required until an anomaly alarm was triggered in order to estimate the average detection time. In terms of TPR, such windows were considered true positives, whereas windows without triggered alarms were considered false negatives. 

\section{Experimental Results}
\label{sect:eval-intrusion-detection}
\subsection{Accuracy}

\begin{figure}
	\includegraphics[width=\columnwidth]{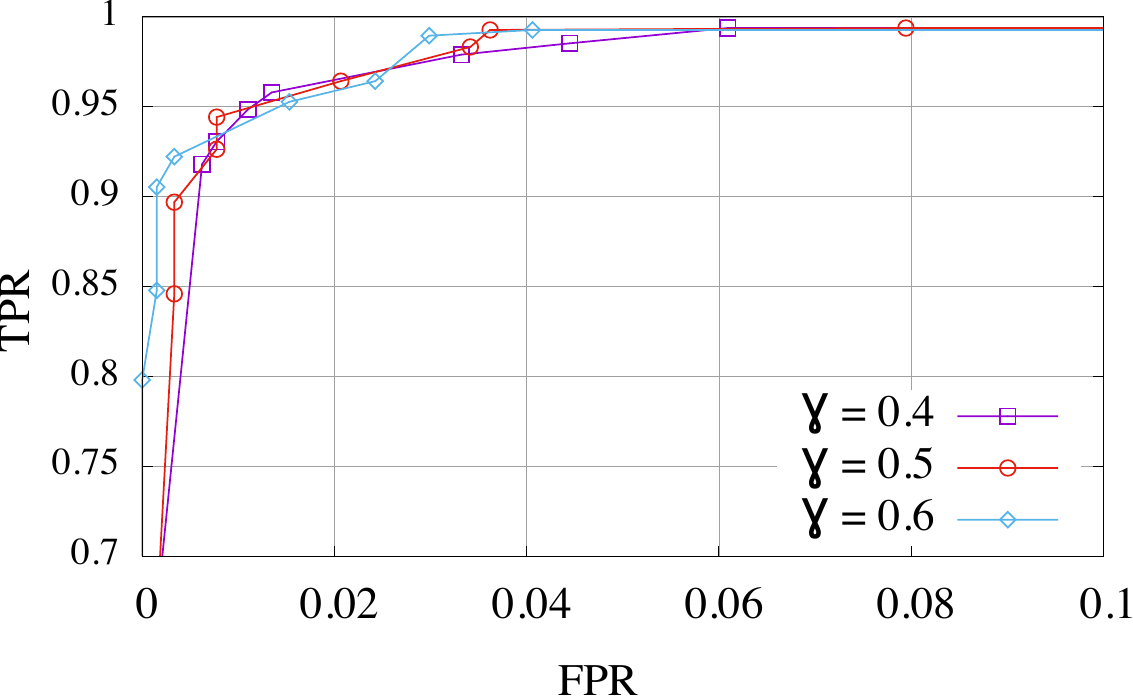}
	\caption{ROC curve of TPR and FPR in dependence of detection threshold $\detThr$ and anomaly triggering threshold $\anomalyThr$.}
	\label{fig:roc-anomaly-detection}
\end{figure}
To determine appropriate values for the detection threshold $\detThr$ and anomaly triggering threshold $\anomalyThr$, we evaluated FPR using the \Activity \textit{(33 devices, 23 device types)} dataset  and TPR using the \Attack \textit{(5 devices, 3 device types)} dataset for a fixed window size of $\win=250$. Fig.~\ref{fig:roc-anomaly-detection} shows the receiver operating characteristic (ROC) curve of FPR and TPR in dependence of these parameters. We can see that all curves quickly reach over 0.9 TPR while keeping a very low FPR (\textless 0.01), which is one of the main objectives for our approach. We therefore select $\detThr = 0.01$ and $\anomalyThr = 0.5$ at $\win = 250$, which achieves $94.01\%$ TPR at \textless 0.01 FPR.

Using these selected parameters in the \Deployment \textit{(14 devices, 10 device types)} dataset and \Attack \textit{(5 devices)} dataset, we achieved an attack detection rate of $95.60\%$ TPR and \textbf{no false positives}, i.e., $0\%$ FPR during one week of evaluation.
These results show that \ourname can successfully address challenge \ChallengeHeterogeneity, \emph{reporting no false alarms in a real-world deployment setting}. Tab.~\ref{tab:attack-detection-times} shows the detailed performance of our system for different attack scenarios (cf. Sect.~\ref{sect:experimental-setup}).
The time to detect attacks varies according to the traffic intensity of the attacks. The average detection delay over all tested attacks is $257\pm 194$~\si{\milli\second}. \ourname can detect an attack in the pre-infection stage after 223 packets while Mirai generates more than 900 packets during pre-infection. It means \ourname is able to detect the attack \emph{even before the attack proceeds to the infection stage}.

The detection rate for DoS attacks is lower than for other attack stages. However, \textit{all DoS attacks are eventually detected} because DoS attacks have a high throughput ($1,412.94$ packets/s.) and we analyze five windows of 250 packets per second at this rate. Considering the $88.96\%$ TPR we achieve on DoS attacks, four windows out of five are detected as anomalous and trigger an alarm. 
It is also worth noting that infected devices in \textit{standby} mode get detected in \textit{33.33\%} of cases, while this activity is very stealthy ($0.05$ packets/s). 


\begin{table}         
	\centering      
	\caption{Average detection times of analyzed Mirai attacks}
	\label{tab:attack-detection-times}
	\begin{tabular}{lrrrl}
		Attack                 & packets/s. & det. time (\SI{}{ms}.) & TPR     \\ 
		\hline
		\textit{standby}       & \textit{0.05}&   \textit{4,051,889} & \textit{33.33\%} &  \\
		\texttt{Pre-Infection} &      426.66  &        524 &   100.00\% &  \\
		\texttt{Infection}     &      721.18  &        272 &    93.45\% &  \\
		\texttt{Scanning}      &      752.60  &        166 &   100.00\% &  \\
		\texttt{DoS}           &    1,412.94  &         92 &    88.96\% &  \\
		\texttt{Average}       &      866.88  & 257$\pm$194 &    95.60\% &  \\		
		\hline
	\end{tabular} 
\end{table}



\subsection{Efficiency of Federated Learning}     
\begin{figure}
	\includegraphics[width=\columnwidth]{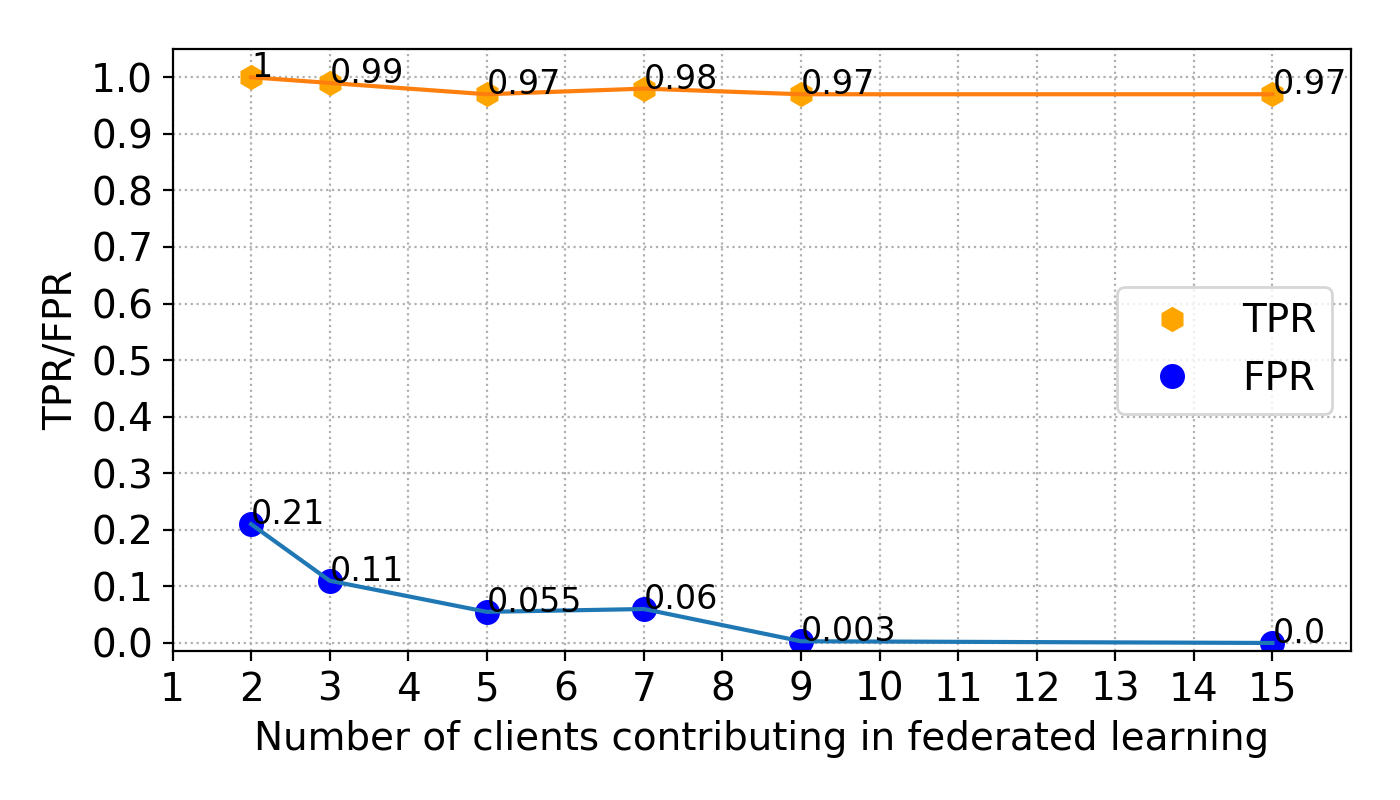}
	\caption{Evolution of TPR and FPR as we increase the number of clients in federated learning. TPR decreases slightly (-3\%) while FPR reaches 0 (-21\%) when using 15 clients.}
	\label{fig:federated}
\end{figure}

We conducted a set of experiments to evaluate federated learning performance with different numbers of clients (ranging from 2 to 15) contributing to the training of the models.
We selected the \textit{number of epochs} that each client trains its local model to be $ 17 $ and specified the \textit{number of communication rounds} between clients and server to be $3$. Therefore, the local models were trained a total of 51 epochs. This was deemed sufficient since in our initial experiments utilizing a centralized learning setting the models converged after approximately 50 epochs. 
%
Each client was allocated a randomized subset of training data from the \Deployment dataset (ranging from 0.1\% to 10\% of the total training dataset size) and we evaluated the system's performance for different numbers of clients involved in building the federated model. \revision{In average, each client takes approximately one second to train one local epoch on its data.} We repeated our experiment three times for each device type, with random re-sampling of the training datasets. As expected, Fig.~\ref{fig:federated} shows that the federated models with more participating clients achieve better FPR, while TPR deteriorates only slightly.

Federated learning provides better privacy for clients contributing to training as they do not need to share their training data. However, this may result in a loss of accuracy of the obtained model in comparison to training the model in a centralized manner. 
To evaluate this possible loss in accuracy, we trained three federated models using the entire training dataset by dividing it among 5, 9 or 15 clients and comparing these to a model trained in a centralized manner. \revision{Tab.~\ref{tab:federated-vs-Centralized} shows a small decrease in TPR as we increase the number of clients while FPR is not deteriorated (remaining constant at $0.00\%$)}.  
This small drop in TPR is not a concern since a large number of packet windows would still trigger an alarm for any attack stage.
 
\begin{table}
	\centering      
	\caption{Effect of using federated learning comparing to centralizing approach}
	\label{tab:federated-vs-Centralized}
\begin{tabular}{|l|c|c|r|r|}
\hline
\multirow{2}{*}{Type} & \multirow{2}{*}{\makecell{Centralized \\learning}} & \multicolumn{3}{c|}{Federated learning}                                                                  \\ \cline{3-5} 
                      &                              & 5 clients                    & \multicolumn{1}{c|}{9 clients} & \multicolumn{1}{c|}{15 clients} \\ \hline
FPR                   & \multicolumn{1}{r|}{0.00\%}     & \multicolumn{1}{r|}{0.00\%}     & 0.00\%                            & 0.00\%                             \\ \hline
TPR                   & \multicolumn{1}{r|}{95.60\%} & \multicolumn{1}{r|}{95.43\%} & 95.01\%                        & 94.07\%                         \\ \hline
\end{tabular}
\end{table}

\subsection{Data Needed for Training}
\label{sect:data-needed-for-training}     
\begin{figure}
	\includegraphics[width=\columnwidth]{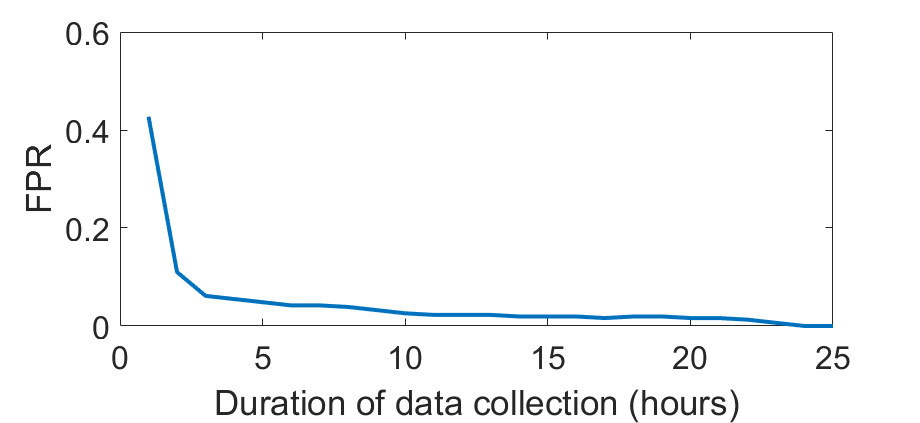}
	\caption{Effect of training data size in time to FPR}
	\label{fig:data-collection-time}
\end{figure}

Fig.~\ref{fig:data-collection-time} shows an example of detection model performance for two Edimax smart plug devices (models 1101W \& 2101W) in dependency of the amount of  data used for training the model. We divided the 7-day \Deployment dataset into one-hour data chunks and randomly sampled different amounts of data chunks for training the model, gradually increasing the training dataset size. The figure shows that the FPR decreases noticeably when the training dataset grows. More importantly, the model needs less than 25 hours of data to achieve FPR~=~0. It shows that our detection model needs little data for training and it means \ourname can address challenge \ChallengeScarcity. Moreover, with the help of our federated learning approach leveraging several clients contributing to training the model, each client needs only a small amount of data i.e, 2.5 hours if there are ten clients involved. It justifies our assumption \textsf{A1} as mentioned in Sect. ~\ref{sect:adversary-model}.

\subsection{Efficiency of Device-Type-Specific Models and Scalability}  
\label{sec:device-typ-specific}
Traditional anomaly detection approaches utilizing a single model for modeling benign behavior easily suffer from increasing false positive rates or decreasing sensitivity when the number of different types of behaviors (i.e., device types) captured by the model grows. This makes them unsuitable for real-world deployments with hundreds or thousands of different device types. Our solution, however, does not have this drawback, as it uses a dedicated detection model for each device type (details in Sect.~\ref{sect:anomaly_detection}). Each of these models focuses solely on the characteristic behavior of one single device type, resulting in more specific and accurate behavioral models, independent of the number of different device types handled by the system. 
To evaluate the benefit of using device-type-specific anomaly detection models compared to using a single model for all devices, we evaluated a single model on the whole \Deployment dataset using 4-fold-cross validation and evaluated detection accuracy on the \Attack dataset. The result is as expected: FPR increases from 0\% to 0.67\% while TPR increases from 95.6\% to 97.21\%. However, as mentioned in Sect.~\ref{sect:system-description}, a high false alarm rate would make the anomaly detection system impractical. If the system had FPR of 0.67\% in our deployment setup, it would trigger around eight alarms per day. It means a smarthome with dozens of devices could have hundreds of false alarms per day.

\subsection{Performance}
We evaluated the processing performance of GRU without specific performance optimizations on a laptop and a desktop computer. The laptop ran Ubuntu Linux 16.04 with an Intel\textcopyright Core\texttrademark i7-4600 CPU with 8GB of memory, whereas the desktop ran Ubuntu Linux 18.04 with an Intel\textcopyright Core\texttrademark i7-7700 CPU with 8GB of memory and a Radeon RX 460 core graphic card. 
We evaluated the processing performance of GRU without specific optimizations on a laptop and a desktop computer. The laptop ran Ubuntu Linux 16.04 with an Intel i7-4600 CPU with 8GB of memory, whereas the desktop ran Ubuntu Linux 18.04 with an Intel i7-7700 CPU with 8GB of memory and a Radeon RX 460 core graphics card with GPU. 
Average processing time per symbol (packet) for prediction was $0.081 (\pm 0.001) $~\si{\milli\second} for the desktop utilizing its GPU and $0.592 (\pm 0.001)$~\si{\milli\second} when executed on the laptop with CPU.
On average, training a GRU model for one device type took 26 minutes on the desktop and 71 minutes on the laptop hardware when considering a week's worth of data in the \Deployment dataset. 
We conclude from this that model training will be feasible to realize in real deployment scenarios, as training will in any case be done gradually as data are collected from the network over longer periods of time.

\section{Effectiveness}
\label{sect:discussion}


\subsection{Generalizability of Anomaly Detection}
\label{sect:generalizability}

 Although we focused our evaluation on the most well-known IoT malware so far: Mirai~\cite{AntonakakisUsenix2017} for the use case, \ourname is likely effective also in detecting other botnet malware like Persirai~\cite{Yeh2017}, Hajime~\cite{Edwards2016}, etc. {\ourname}'s anomaly detection leverages deviations in the behavior of infected IoT devices caused by the malware. Such deviations will be observable for any malware.


\subsection{Evolution of IoT Device Behavior}
\label{sect:evolution}
The behavior of an IoT device type can evolve due to, e.g., firmware updates that bring new functionality. This modifies its behavior and may trigger false alarms for legitimate communication. We prevent these false alarms by correlating anomaly reports from all {\sgw}s at the \iotss. Assuming firmware updates would be propagated to many client networks in a short time, if alarms are reported from a large number of security gateways for the same device type in a short time, we can cancel the alarm and trigger re-learning of the corresponding device identification and anomaly detection models to adapt to this new behavior. 
To ensure that sudden widespread outbreaks of an IoT malware infection campaign are not erroneously interpreted as firmware updates, the canceling of an alarm can be confirmed by a human expert at the \iotss. This should represent a small burden, as the roll-out of firmware updates is a relatively seldom event.

\subsection{Mimicking Legitimate Communication}
\label{sect:Mimicking}
An adversary that has compromised an IoT device can attempt to mimic the device's legitimate communication patterns to try to remain undetected. However, as the device-type-specific detection model is restricted to the (relatively limited) functionality of the IoT device, it is in practice very difficult for the adversary to mimic legitimate communication and at the same time achieve a malicious purpose, e.g., scanning, flooding, etc., especially when considering that any change in packet flow semantics is also likely to change the characteristics (protocol, packet size, port, etc.) of packets and their ordering, which are both used for detecting anomalies in the packet sequence.
Moreover, adversaries would need to know the device-type-specific communication patterns in order to mimic them. This makes it significantly harder for adversaries to develop large scale IoT malware that affects a wide range of different IoT device types in the way that, e.g., Mirai does.

\subsection{Adversarial Machine Learning}
\label{sect:adversarial}
\noindent\textbf{Adversarial examples.} If an adversary manages to compromise an IoT device while remaining undetected, it can attempt to 'poison' the training process of the system by forging packets as \emph{adversarial examples} that are specifically meant to influence the learning of the model in such a way that malicious activities are not detected by it. There exist techniques to forge adversarial examples to neural networks~\cite{carlini2017towards}. However, these apply to images~\cite{kurakin2016adversarial,evtimov2017robust} and audio inputs~\cite{vaidya2015cocaine,zhang2017dolphinattack}, where objective functions for the underlying optimization problem are easy to define. 
Forging adversarial examples consists of finding minimal modifications $\epsilon$ for an input $x$ of class $C$ such that $x + \epsilon$ is classified as $C' \neq C$. For example, in our case this would mean that a malicious packet is incorrectly classified as a benign one. In contrast to image or audio processing, however, our features (symbols) are not raw but processed from packet properties. First, it means that modifications $\epsilon$ are computed for our symbolic representation of packet sequences which are difficult to realize in a way that would preserve their utility for the adversary, i.e., realize 'useful' adversarial functionality required for malicious activities like scanning or DoS. Second, it is difficult to define the objective distance to minimize in order to achieve ``small modifications'' since modifying the value of one single packet characteristic (protocol, port, etc.) can change the semantics of a packet entirely.

\noindent\textbf{Poisoning federated learning.}
For initial model training, we can assume the training data contain only legitimate network traffic, as devices are assumed  initially to be benign (assumption \textsf{A1} (Sect.~\ref{sect:adversary-model})). 
However, the federated setting can be subject to \emph{poisoning attacks} during re-training, where the adversary uses adversarial examples as described above to corrupt the anomaly detection model so that it eventually will accept malicious traffic as benign~\cite{biggio2012poisoning} (or vice versa).
Techniques have been developed for preventing poisoning attacks by using local outlier detection-based filtering of adversarial examples to pre-empt model poisoning~\cite{rubinstein2009antidote}. 


%
In the scope of this paper we assume that the \sgw is not compromised by the adversary (assumption \textsf{A2} (Sect.~\ref{sect:adversary-model})). However, since a malicious user can have physical access to his \sgw, it is thinkable that he could compromise it in order to stage a poisoning attack against the system using adversarial examples. In this case, local filtering of adversarial examples is not possible, as it can not be enforced by the compromised \sgw. We are therefore currently focusing our ongoing research efforts on applying poisoning mitigation approaches applied at the \iotss. These include using more robust learning approaches less resilient to adversarial examples that will 'average out' the effects of adversarial examples, as well as approaches similar to, e.g.,  Shen \emph{et al.}~\cite{Shen:2016:UDA:2991079.2991125}, where malicious model updates are detected before they are incorporated in global detection models.

\section{Related Work}
\label{sect:relatedwork}

Several solutions have been proposed for the detection and prevention of intrusions in IoT networks~\cite{JiaNDSS2017-ContexIOT,Raza2013svelte}, sensor networks~\cite{rajasegarar2014hyperspherical} and industrial control systems~\cite{Jardine:2016:SSN:2994487.2994496,Kleinmann:2016:ACS:2994487.2994490}.
SVELTE~\cite{Raza2013svelte} is an intrusion detection system for protecting IoT networks from already known attacks. It adapts existing intrusion detection techniques to IoT-specific protocols, e.g., 6LoWPAN. 
Similarly, Doshi at al. ~\cite{Doshi2018DDoSDetection} proposed a signature-based approach to detect known DDoS attacks using features representing the density of the network traffic.
In contrast, \ourname performs dynamic detection of any unknown attacks that deviate from the legitimate behavior of the device, since it only models legitimate network traffic.
Jia \emph{et al.}~\cite{JiaNDSS2017-ContexIOT} proposed a context-based system to automatically detect sensitive actions in IoT platforms. 
This system is designed for patching vulnerabilities in appified IoT platforms such as Samsung SmartThings. It is not applicable to multi-vendor IoT systems while \ourname is platform independent.


Detecting anomalies in network traffic has a long history~\cite{krugel2002service,portnoy2001intrusion,sekar2002specification,sommer2010outside}. Existing approaches rely on analysing single network packets~\cite{krugel2002service} or clustering large numbers of packets~\cite{portnoy2001intrusion,rajasegarar2014hyperspherical} to detected intrusions or compromised services. 
Some works have proposed, as we do, to model communication as a language~\cite{Kleinmann:2016:ACS:2994487.2994490,sekar2002specification}. For instance, authors of~\cite{sekar2002specification} derive finite state automatons from layer 3-4 communication protocol specifications. Monitored packets are processed by the automaton to detect deviations from protocol specification or abnormally high usage of specific transitions.
Automatons can only model short sequences of packets while we use GRU to model longer sequences, which enables the detection of stealthy attacks. Also, modelling protocol specifications is coarse and leaves room for circumventing detection. In contrast, we use finer grained-features that are difficult to forge while preserving the adversarial utility of malicious packets.


Lately, recurrent neural networks (RNN) have been used for several anomaly-detection purposes.
Most applications leverage Long Short-Term Memory (LSTM) networks for detecting anomalies in time series~\cite{malhotra2015long}, aircraft data~\cite{nanduri2016anomaly} or system logs~\cite{Du:2017:DAD:3133956.3134015}.
Oprea at al.~\cite{oprea2015detection} use deep belief networks for mining DNS log data and detect infections in enterprise networks. In contrast to these works, \ourname uses a different flavor of RNN, namely GRU that can be learned using less training data, enabling \ourname to be trained faster,
and operate in real-time, detecting anomalies in live network traffic instead of utilizing off-line analysis.

\section{Summary}
In this paper we introduced {\ourname}: a self-learning system for detecting compromised devices in IoT networks.
Our solution relies on novel automated techniques for \textit{device-type-specific anomaly detection}. 
\ourname does \textit{not require any human intervention or labeled data} to operate. It learns anomaly detection models autonomously, using unlabeled  crowdsourced data captured in client IoT networks.
We demonstrated the efficacy of anomaly detection in detecting a large set of malicious behavior from devices infected by the Mirai malware.
\ourname detected 95.6\% of attacks in 257 milliseconds on average and without raising any false alarm when evaluated in a real-world deployment.


\noindent\textbf{Acknowledgments}: \revision{This work was supported in part by the Academy of Finland (SELIoT project - grant 309994), the German Research Foundation (DFG) within CRC 1119 CROSSING (S2 and P3) and the Intel Collaborative Institute for Collaborative Autonomous and Resilient Systems (ICRICARS). We would also like to thank Cisco Systems, Inc. for
their support of this work.}

\bibliographystyle{IEEEtran}

\end{document}